\documentclass[twocolumn,floatfix,showpacs,showkeys,preprintnumbers,nofootinbib,superscriptaddress]{revtex4}
\usepackage[utf8]{inputenc}
\usepackage[sort&compress]{natbib}
\usepackage{ulem}
\usepackage{bm}
\usepackage{times}
\usepackage{amssymb,amsbsy,amsmath,amsfonts}
\usepackage{graphicx}
\usepackage{float}
\usepackage{color}
\usepackage{morefloats}
\usepackage{rotating}
\usepackage{srcltx}
\usepackage{slashed}
\usepackage{subfigure}
\usepackage{multirow}
\usepackage{verbatim}
\usepackage{hyperref}
\usepackage{tabularx}

\begin{document}

\title{$X_0(2866)$ as a $D^*\bar{K}^*$ molecular state}

\author{Ming-Zhu Liu}
\affiliation{School of Physics, Beihang University, Beijing 100191, China}

\author{Jun-Jun Xie}\email{xiejujun@impcas.ac.cn}
\affiliation{Institute of Modern Physics, Chinese Academy of Sciences, Lanzhou 730000, China}
\affiliation{School of Nuclear Sciences and Technology, University of Chinese Academy of Sciences, Beijing 101408, China}
\affiliation{School of Physics and Microelectronics, Zhengzhou University, Zhengzhou, Henan 450001, China}

\author{Li-Sheng Geng}\email{lisheng.geng@buaa.edu.cn}
\affiliation{School of Physics, Beihang University, Beijing 100191, China}
\affiliation{Beijing Key Laboratory of Advanced Nuclear Materials and Physics, Beihang University, Beijing 100191, China}
\affiliation{School of Physics and Microelectronics, Zhengzhou University, Zhengzhou, Henan 450001, China}
\affiliation{Beijing Advanced Innovation Center for Big Data-Based Precision Medicine, School of Medicine and Engineering, Beihang University, Beijing 100191, China}

\date{\today}

\begin{abstract}
Very recently the LHCb Collaboration reported the discovery of two open charm tetraquark states, $X_{0}(2866)$ and $X_{1}(2904)$.  In the present work,  we study the $D^{(\ast)}$ and $\bar{K}^{(\ast)}$ interaction in the one-boson exchange model and show that the $X_{0}(2866)$ can be understood as a $D^{\ast}\bar{K}^{\ast}$  molecule   with $I(J^{P})=0(0^{+})$, or at least it has a large molecular component. On the other hand,   the $X_{1}(2904)$ can not be interpreted as a molecular state.  Inspired by the discovery of the $X_0(2866)$ and the fact that the $D^*\bar{K}^*$ interaction is strong enough to generate a bound state, we also discuss likely existence of other open charm molecules. In the meson-meson sector, two molecules near the mass thresholds of $DD^{\ast}$ and $D^{\ast}D^{\ast}$ with $I(J^{P})=0(1^{+})$ are obtained,  and  using the heavy quark flavor symmetry their $\bar{B}\bar{B}^{\ast}$ and $\bar{B}^{\ast}\bar{B}^{\ast}$ counterparts are also predicted. In the meson-baryon sector, 7 open charm molecules with $I=1/2$ near the mass thresholds of $D^{(\ast)}\Sigma_{c}^{(\ast)}$ naturally appear, as dictated by the heavy quark spin symmetry.
\end{abstract}


\maketitle

\section{Introduction}
Since the Belle Collaboration discovered the $X(3872)$~\cite{Choi:2003ue},  a lot of exotic states that can not easily fit into the conventional quark model have been reported by collaborations all over the world, which not only offers the chance to study the strong interaction, but also challenges the existing theories and models.  There are different interpretations of these states, such as molecular states, compact multiquark states, traditional hadrons, and kinematics effects. Among them hadronic molecules are a popular explanation, which were conjuncted four decades  ago from a direct analogy to the deuteron and the nuclear force that binds it~\cite{Voloshin:1976ap,DeRujula:1976zlg}.
After extensive studies on the production, decay, and mass spectra of these exotic states, some were regarded as robust molecular candidates, such as the $X(3872)$, the $P_{c}(4312)$, the $P_{c}(4440)$, and the $P_{c}(4457)~$\cite{Chen:2016qju,Sun:2011uh,Liu:2008tn,Liu:2019stu,Chen:2019asm,Liu:2019tjn,Xiao:2019aya,Xiao:2019mvs,Lin:2019qiv,Yamaguchi:2019seo,Liu:2019zvb,Valderrama:2019chc,Du:2019pij}.   The particular property of  these molecular candidates is that they carry hidden charm number.  An intriguing question is then whether open charm molecular candidates exist.

In 2003, the BaBar Collaboration reported a narrow structure in the $D_{s}^{+}\pi^{0}$ invariant mass distribution with a mass of 2.32 GeV  and quantum number $J^{P}=1^{+}$~\cite{Aubert:2003fg}, whose mass is much lower than the quark model prediction~\cite{Godfrey:1985xj}. To solve this problem,  a molecule composed of $DK$ and $D_{s}\eta$   was  proposed by many theoreticians~\cite{Barnes:2003dj,Gamermann:2006nm,Guo:2006fu,Faessler:2007gv,Liu:2012zya,Altenbuchinger:2013vwa}.   Moreover,  its heavy quark spin partner $D_{s1}(2460)$ can also be described as a $D^{\ast}K$ molecule~\cite{Faessler:2007us,Altenbuchinger:2013vwa}.  Thus  $D_{s0}(2317)$ and  $D_{s1}(2460)$ can be treated as hadronic molecular candidates with open charm.  In the molecular picture their quark contents are $c\bar{s}q\bar{q}$.  If the quark contents are changed to $c{s}\bar{q}\bar{q}$, which indicate $D\bar{K}$ at the hadron level, whether such hadronic molecules exist or not is of great interests both theoretically and experimentally.

 Very recently  the LHCb Collaboration reported the discovery of two new states, $X_{0}(2866)$ and $X_{1}(2904)$, in the $D^{+}K^{-}$ invariant mass spectrum   of the $B^{\pm}\rightarrow D^{+}D^{-}K^{\pm}$ decay~\cite{lhcb-x2900}.    Their corresponding quantum number, mass, and width are:
\begin{eqnarray}
J^{P}&=&0^{+}:  \quad   m=2866\pm 7,  \quad  \Gamma=57\pm 13, \\
J^{P}&=&1^{-}:  \quad   m=2904\pm 5,  \quad  \Gamma=110\pm 12,
\end{eqnarray}
all in units of MeV, whose statistical significance is more than 5$\sigma$. The quark contents of these two states are $c{s}\bar{q}\bar{q}$,  who are the first open charm tetraquark states discovered experimentally so far. In Ref.~\cite{Karliner:2020vsi} using two kinds of quark model Karliner and Rosner  identified the $X_{0}(2866)$ as a $cs\bar{u}\bar{d}$ tetraquark, while the $X_{1}(2904)$ can not be described as a tetraquark state. Since both states are located close to the mass threshold of $D^{\ast}\bar{K}^{\ast}$, a natural question is whether they
are $D^*\bar{K}^*$ molecules.  In this work we will investigate whether they can be interpreted as $D^{\ast}\bar{K}^{\ast}$ bound states within the one-boson exchange (OBE) model.

This paper is organized as follows. We briefly explain the pertinent ingredients of the OBE model in section II. We show in Section III that it is possible that the $X_0(2866)$ has a large molecular component, while we cannot generate a $2^+$ state. Inspired by such possibilities, we further study a number of related systems, including $D^{(\ast)}D^{(\ast)}$, $\bar{B}^{(\ast)}\bar{B}^{(\ast)}$,  and  $D^{(\ast)}\Sigma_{c}^{(\ast)}$, and find some molecule candidates. We conclude in the last section.

\section{Formulism}
\label{sec:obe}

We will explore whether the ${D}^{(\ast)}\bar{K}^{(\ast)}$
interactions are strong enough to form bound states. The relevant
Lagrangians are the same as those of the
${D}^{(\ast)}D^{(\ast)}$ interactions~\cite{Liu:2019stu}.     The Lagrangian describing the interaction
between a charmed and a light meson can be written as
\begin{eqnarray}
    \mathcal{L}_{H H \pi} &=& -\frac{g}{\sqrt{2} f_{\pi}}\,{\rm Tr}
  \left[ {H}^{\dagger} \vec{\sigma} \cdot \nabla ( \vec{\tau} \cdot \vec{\pi})
    H \right] \, , \label{eq:L-pi} \\
  \mathcal{L}_{H H \sigma} &=& g_{\sigma}\,{\rm Tr}\left[
    {H}^{\dagger} \sigma H \right] \, , \label{eq:L-sigma} \\
  \mathcal{L}_{H H \rho} &=& g_{\rho}\,{\rm Tr}\left[
    {H}^{\dagger} \vec{\tau} \cdot \vec{\rho}^{0} H \right]
  \nonumber \\
  &-& \frac{f_{\rho 1}}{4 M}\,\epsilon_{ijk}\,{\rm Tr}\left[
    {H}^{\dagger} \sigma_k \vec{\tau} \cdot \left( \partial_i \vec{\rho}_j
    - \partial_j \vec{\rho}_i \right) H \right] \, , \label{eq:L-rho} \\
    \mathcal{L}_{H H \omega} &=& -g_{\omega}\,{\rm Tr}\left[
    {H}^{\dagger} {\omega}^{0} H \right]
  \nonumber \\
  &+& \frac{f_{\omega}}{4 M}\,\epsilon_{ijk}\,{\rm Tr}\left[
    {H}^{\dagger} \sigma_{k} \, \left( \partial_i {\omega}_j
    - \partial_j {\omega}_i \right) H \right] \, , \label{eq:L-omega}
\end{eqnarray}
where ${H} = \frac{1}{\sqrt{2}}( D + \vec{D}^* \cdot \vec{\sigma})$, satisfying the constraint of heavy quark spin
symmetry. The coupling of $\pi$ to ${D}^{(\ast)}$ are  $g=0.6$ and $f_{\pi}=0.132$ GeV.  The sigma coupling is determined from the nucleon-nucleon-sigma
coupling in the non-linear sigma model
($g_{\sigma NN} = \sqrt{2}\,M_N / f_{\pi} \simeq 10.1$)
and the quark model relation:
$ g_{\sigma} = \frac{1}{3} g_{\sigma NN} \simeq 3.4$. The $\rho$ and $\omega$ couplings to
$D^{(\ast)}$ are of both electric-type ($g_{v^{\ast}}$) and
magnetic-type  ($f_{v^{\ast}}$).   We obtain
$g_{v}=2.6$ and  $f_{v}=4.5$ using a
common mass $M=1.867$ GeV for the normalization~\cite{Liu:2019stu}. Here SU(2) symmetry is adopted in our calculation, regarding strange and charm quarks as spectators. Thus in our work  the Lagrangian describing the interaction between the $\bar{K}^{(\ast)}$ and a light meson are the same as that describing the interaction between the $D^{(\ast)}$ and a light meson.

For the ${D}\bar{K}$, $D^{\ast}\bar{ K}$, and $D \bar{K}^{\ast}$ systems  the exchanged light mesons can  be $\sigma$, $\rho$, or $\omega$. While $\pi$, $\sigma$, $\rho$, and $\omega$ light mesons are allowed  for the ${D}^{\ast} \bar{K}^{\ast}$ system.  From the Lagrangians of Eq.~(\ref{eq:L-rho}) we can derive the OBE potentials
for the ${D}^{(\ast)} \bar{K}^{(\ast)}$ systems as
\begin{eqnarray}
  V_{\pi}(\vec{r}) &=&
  \vec{\tau}_{1} \cdot \vec{\tau}_{2}\,\frac{g^{2}}{6 f_{\pi}^2}\,\Big[
    - \vec{\sigma}_{1} \cdot \vec{\sigma}_{2}\,\delta(\vec{r})
    \nonumber \\ && \quad
    + \, \vec{\sigma}_{1} \cdot \vec{\sigma}_{2}\,m_{\pi}^3\,W_Y(\mu_{\pi} r)
    \nonumber \\ && \quad
    + \, S_{12}(\vec{r})\,m_{\pi}^3\,W_T(m_{\pi} r) \Big] \, , \label{11}  \\
  V_{\sigma}(\vec{r}) &=& -{g_{\sigma }^2}\,m_{\sigma}\,W_Y(m_{\sigma} r)
  \, ,  \label{22} \\
  V_{\rho}(\vec{r}) &=&\vec{\tau}_{1} \cdot \vec{\tau}_{2}\,\Big[
    {g_{v}^2}\,m_{\rho}\,W_Y(m_{\rho} r) \nonumber \\
    && \quad + \frac{f_{v}^2}{4 M^2}\,\Big(
    -\frac{2}{3}\,\vec{\sigma}_{1} \cdot \vec{\sigma}_{2} \delta(\vec{r})
    \nonumber \\ && \quad
    +\frac{2}{3}\,\vec{\sigma}_{1} \cdot \vec{\sigma}_{2}
    \, m_{\rho}^3 \, W_Y(m_{\rho} r)
    \nonumber \\ && \quad
    -\frac{1}{3}\,S_{12}(\hat{r})\, m_{\rho}^3 \, W_T(m_{\rho} r) \,\,
    \Big) \, \Big] \, , \label{33}
    \end{eqnarray}
\begin{eqnarray}
    V_{\omega}(\vec{r}) &=&
    {g_{v}^2}\,m_{\omega}\,W_Y(m_{\omega} r) \nonumber \\
    && + \frac{f_{v}^2}{4 M^2}\,\,\Big[
    -\frac{2}{3}\,\vec{\sigma}_{1} \cdot \vec{\sigma}_{2} \, \delta(\vec{r})
    \nonumber \\ &&
    +\frac{2}{3}\,\vec{\sigma}_{1} \cdot \vec{\sigma}_{2} \,
    m_{\omega}^3 \, W_Y(m_{\omega} r)
    \nonumber \\ &&
    -\frac{1}{3}\,S_{12}(\hat{r})\, m_{\omega}^3 \, W_T(\mu_{\omega} r) \,\,
    \Big] \, ,  \label{44}
\end{eqnarray}
where the dimensionless functions $W_Y(x)$ and $W_T(x)$ are defined as
\begin{eqnarray}
  W_Y(x) &=& \frac{e^{-x}}{4\pi x} \, , \\
  W_T(x) &=& \left( 1 + \frac{3}{x} + \frac{3}{x^2} \right)
  \,\frac{e^{-x}}{4\pi x} \, .
\end{eqnarray}

As the charmed mesons involved in our calculation are not point-like
particles, we introduce a form factor to take into account the finite
sizes of charmed mesons. Here we  use a monopolar form factor (for
more details we refer to  Refs.~\cite{Liu:2019stu,Liu:2019zvb})
\begin{eqnarray}
  F(q, m, \Lambda)=\frac{\Lambda^{2}-m^2}{\Lambda^{2}-{q}^2} \label{Eq:FF} \,
  ,
\end{eqnarray}
where $m$ and $q$ are the mass and 4-momentum of the exchanged
meson, respectively.

Using the above form factor, the functions $\delta$, $W_{Y}$, and
$W_{T}$ in Eqs.~(\ref{11}-\ref{44}) need to changed accordingly
\begin{eqnarray}
  \delta(r) &\to& m^3\,d(x,\lambda) \, , \\
  W_Y(x) &\to& W_Y(x, \lambda) \, , \\
  W_T(x) &\to& W_T(x, \lambda) \, ,
\end{eqnarray}
with $\lambda = \Lambda / m$.
The corresponding functions $d$, $W_Y$, and $W_T$ read
\begin{eqnarray}
  d(x, \lambda) &=& \frac{(\lambda^2 - 1)^2}{2 \lambda}\,
  \frac{e^{-\lambda x}}{4 \pi} \, , \\
  W_Y(x, \lambda) &=& W_Y(x) - \lambda W_Y(\lambda x) \nonumber \\ && -
  \frac{(\lambda^2 - 1)}{2 \lambda}\,\frac{e^{-\lambda x}}{4 \pi} \, , \\
  W_T(x, \lambda) &=& W_T(x) - \lambda^3 W_T(\lambda x) \nonumber \\ && -
  \frac{(\lambda^2 - 1)}{2 \lambda}\,\lambda^2\,
  \left(1 + \frac{1}{\lambda x} \right)\,\frac{e^{-\lambda x}}{4 \pi} \, .
\end{eqnarray}

\begin{table*}[htbp]
\centering \caption{Matrix elements of spin-spin and
  tensor operators for the partial waves considered
  in this work.} \label{tab:tensor}
\begin{tabular}{c|c|c|c|c}
\hline\hline Molecule & Partial Waves & $J^P$ & $\vec{a}_1 \cdot
\vec{a}_2 $ & $S_{12} = 3\,\vec{a}_1 \cdot \hat{r} \, \vec{a}_2
\cdot \hat{r} - \vec{a}_1 \cdot \vec{a}_2$
\\ \hline
${D}^{\ast} \bar{K}^{\ast}$ & $^1S_0$-$^5D_0$  & $0^+$ &
$\left(\begin{matrix}
-2 & 0 \\
0 & 1 \\
\end{matrix}\right)$& $\left(\begin{matrix}
0 & -\sqrt{2} \\
-\sqrt{2} & -2 \\
\end{matrix}\right)$  \\ \hline
${D}^{\ast} \bar{K}^{\ast}$ & $^3S_1$-$^3D_1$ & $1^+$ &
$\left(\begin{matrix}
-1 & 0 \\
0 & -1 \\
\end{matrix}\right)$& $\left(\begin{matrix}
0 & \sqrt{2} \\
\sqrt{2} & -1 \\
\end{matrix}\right)$  \\ \hline
${D}^{\ast} \bar{K}^{\ast}$ & $^1D_2$-$^5S_2$-$^5D_2$& $2^+$ &
$\left(\begin{matrix}
-2 & 0 & 0  \\
 0 & 1 & 0  \\
 0 & 0 & 1  \\
\end{matrix}\right)$& $\left(\begin{matrix}
0 & -\sqrt{\frac{2}{5}} & \frac{2}{\sqrt{7}}  \\
-\sqrt{\frac{2}{5}} & 0 & \sqrt{\frac{14}{5}} \\
\frac{2}{\sqrt{7}} & \sqrt{\frac{14}{5}} &
\frac{3}{7} \\
\end{matrix}\right)$  \\
\hline\hline
\end{tabular}
\end{table*}

In addition, since  $D$-wave interactions may play an important
role in forming hadronic molecules, we also consider the $D$-wave
interaction in the ${D}^{\ast} \bar{K}^{\ast}$ system, which will
produce two  kinds of spin operators, spin-spin
$\vec{\sigma}\cdot\vec{\sigma}$ and tensor $ S_{12}$. We
define the tensor operator as $S_{12}=3\,\vec{\sigma}_1 \cdot \hat{r} \,
\vec{\sigma}_2 \cdot \hat{r} - \vec{\sigma}_1 \cdot \vec{\sigma}_2$.
The specific matrix element of spin-spin and tensor operators for
the ${D}^{\ast} \bar{K}^{\ast}$ system are displayed  in Table
\ref{tab:tensor}.

\section{Numerical results and Discussions}
\label{sec:pre}

\subsection{$cs\bar{q}\bar{q}$}

\begin{figure}[htbp]
 \center{\includegraphics[scale=0.35]  {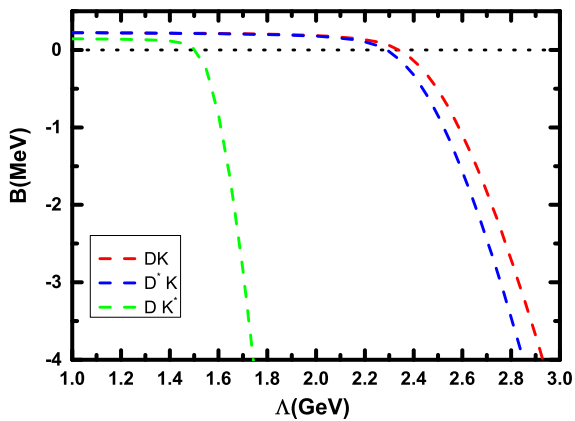}}
 \caption{\label{ddsbar1}Binding energies of $D\bar{K}$, $D^{\ast}\bar{K}$, and $D\bar{K}^{\ast}$ molecules as a function of  the cutoff. }
 \end{figure}

\begin{figure}[htbp]
 \center{\includegraphics[scale=0.32]  {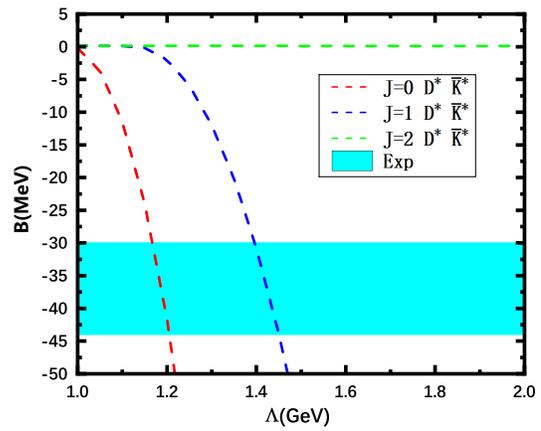}}
 \caption{\label{ddsbar2} Binding energies of ${D}^{\ast}\bar{K}^{\ast}$ ($I=0$) molecules as a function of  the cutoff.  }
 \end{figure}

In Figs.~\ref{ddsbar1} and \ref{ddsbar2}, we show the binding
energies of the ${D}^{(\ast)}\bar{K}^{(\ast)}$ systems as a
function of the cutoff. The OBE potentials of $D\bar{K}$, $D^{\ast}\bar{K}$, and $D\bar{K}^{\ast}$ are the same, and the only difference of these states  are the kinetic energy. We present the binding energy dependence of these states on the cutoff  in Fig.~\ref{ddsbar1}.  The results indicate that  a
bound  state near the mass threshold of ${D}\bar{K}^{\ast}$ is likely to exist, but with a cutoff much larger than the
preferred value of about 1 GeV.  An even larger cutoff is needed for the $D\bar{K}$ and $D^{\ast}\bar{K}$ systems to bind. Therefore we conclude that the existence of  molecular states near the $D\bar{K}$ and $D^{\ast}\bar{K}$ mass thresholds is unlikely.

 For the $D^* \bar{K}^*$ system, there are three spin-parity assignments, $J=0$, $J=1$, and $J=2$, for $S$-wave interactions (which we focus on in the present work).  From the results shown in Fig~\ref{ddsbar2}, it seems that two bound states with $J^{P}=0^{+}$ and $J^{P}=1^{+}$ with $I=0$ are  likely. Therefore the $X_{0}(2866)$ could be assigned as a molecule state near the mass threshold $D^{\ast}\bar{K}^{\ast}$ with $J^{P}=0^{+}$.  In addition, our results do not support the existence of a $J^{P}=2^{+}$ state because the rather large cutoff needed.~\footnote{We did not find any $I=1$ $D^*\bar{K}^*$ bound states with a reasonable cutoff.} 

In our framework, if one considers the $X_{1}(2904)$ as a $D^{\ast}\bar{K}^{\ast}$ bound state, the interaction between $D^*$ and $\bar{K}^*$ is $P$-wave, which results a much more repulsive potential.  Thus the $X_{1}(2904)$ can not be identified as a $D^{\ast}\bar{K}^{\ast}$ molecule in our framework.

Furthermore, we note that the cutoff needed to make the mass of the $X_0(2866)$ to agree with the LHCb data is about 1.2 GeV, slightly larger the preferred value of 1 GeV. As a result, such a relatively large cutoff may indicate the existence of a sizable compact tetraquark component. In this sense, the conclusion of our present study does not necessarily contradict that of Ref.~\cite{Karliner:2020vsi}, at least not strongly.

\subsection{$cc\bar{q}\bar{q}$}
\begin{figure}[htbp]
 \center{\includegraphics[scale=0.35]  {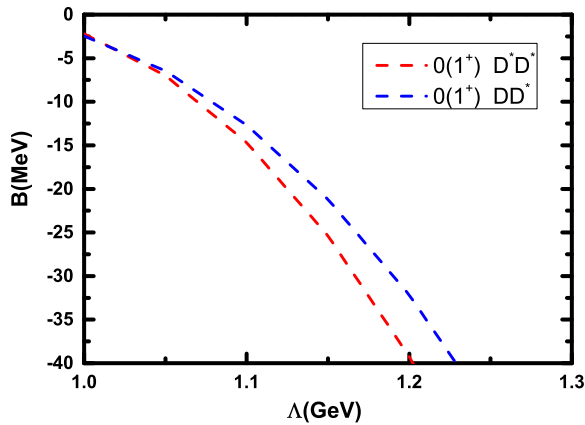}}
 \caption{\label{dk3} Binding energies of ${D}^{\ast} D$ and $D^{\ast}D^{\ast}$ molecules as a function of  the cutoff.  }
 \end{figure}

 If we replace the $s$ quark in the $cs\bar{q}\bar{q}$ system with a $c$ quark, the system becomes $cc\bar{q}\bar{q}$ with charm number $C=2$.
At the hadron level, this system corresponds to $D^{(\ast)}D^{(\ast)}$.  In our previous work~\cite{Liu:2019stu}, assuming that the $X(3872)$ is a $1^{++}$ bound state, we have already investigated the $\bar{D}^{(\ast)}D^{(\ast)}$ system in the OBE model. Two $0(1^{+})$ hadronic molecules with $C=2$ near the $DD^{\ast}$ and  $D^{\ast}D^{\ast}$ mass thresholds are predicted.  In Fig~\ref{dk3}, we display the dependence of the binding energy of the two systems on the cutoff. It seems to be very likely that two hadronic molecules close to the $DD^{\ast}$ and  $D^{\ast}D^{\ast}$ thresholds exist, which should be clarified by future experiments.

\subsection{$bb\bar{q}\bar{q}$}
\begin{figure}[htbp]
 \center{\includegraphics[scale=0.35]  {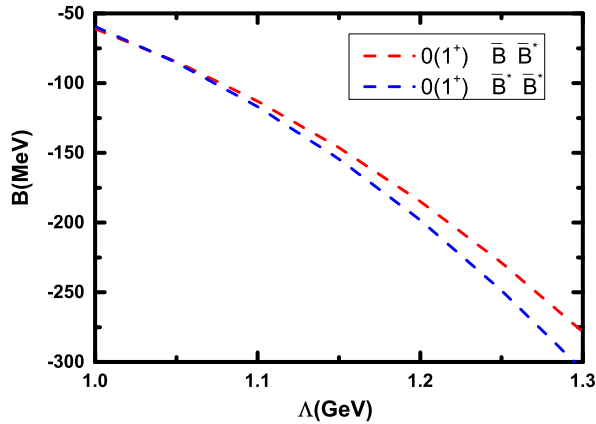}}
 \caption{\label{dk4} Binding energies of $\bar{B}^{\ast} B$ and $\bar{B}^{\ast}\bar{B}^{\ast}$ molecules as a function of the cutoff.  }
 \end{figure}

Using the heavy quark flavor symmetry, one can straightforwardly correlate   the $\bar{B}^{(\ast)}\bar{B}^{(\ast)}$ system with 
the $D^{(\ast)}D^{(\ast)}$ system.
Since the potential in the bottom sector is the same as that  in the charm sector, the binding energy is larger than that in the charm sector as shown in Fig~\ref{dk4}. Thus deeper bound states are more likely to exist in the open bottom sector.

\subsection{$cc{q}q\bar{q}$}

In  our previous studies, we have already confirmed that there is a complete multiplet of hadronic molecules constrained by heavy quark spin symmetry in the $\bar{D}^{(\ast)}\Sigma_{c}^{(\ast)}$ system in light of the LHCb pentaquark states~\cite{Liu:2019stu,Liu:2019zvb}, but  these hadronic molecules have a hidden charm number.   Here, we further explore whether there exist open charm molecules in the $D^{(\ast)}\Sigma_{c}^{(\ast)}$ system.  In terms of $G$-party~\cite{Klempt:2002ap}, the OBE potential of  $D^{(\ast)}\Sigma_{c}^{(\ast)}$ can be easily derived from that of   $\bar{D}^{(\ast)}\Sigma_{c}^{(\ast)}$.  In this system the cutoff should not be treated as a free parameter because the cutoff can be determined by reproducing the binding energy of the $P_{c}(4312)$~\cite{Liu:2019zvb}. Within the OBE model and with $\Lambda=1.119$ GeV, one can study the $D^{(\ast)}\Sigma_{c}^{(\ast)}$ systems. The results are displayed in Table~\ref{dk5}, which show the existence of 7 open charm molecules made of  $D^{(\ast)}\Sigma_{c}^{(\ast)}$. We further note  that the binding energies of the $D^{(\ast)}\Sigma_{c}^{(\ast)}$  states are larger than those of the $\bar{D}^{(\ast)}\Sigma_{c}^{(\ast)}$,  because the $\pi$ and $\omega$ OBE potentials change sign in the former case compared to those in the latter case.
\begin{table}[ttt]
\centering
\caption{Binding energies of ${D}^{(\ast)}\Sigma_{c}^{(\ast)}$ molecules with $I=1/2$ }
\label{dk5}
\begin{tabular}{cccccc}
\hline\hline
Molecule  & $I$ & $J^{P}$  & B.E (MeV) &  Mass (MeV) \\
\hline
${D}\Sigma_c$ & $\tfrac{1}{2}$ & $\tfrac{1}{2}^-$ &
$31.7^{+16.6}_{-13.9}$  &4289.3 \\
\hline
${D}\Sigma_c^*$ & $\tfrac{1}{2}$ & $\tfrac{3}{2}^-$ &
$32.5^{+16.8}_{-14.1}$ & $4352.5$ \\
\hline
${D}^*\Sigma_c$ & $\tfrac{1}{2}$ & $\tfrac{1}{2}^-$
&  $18.4^{+11.9}_{-9.3}$  & $4444.6$  \\
${D}^*\Sigma_c$ & $\tfrac{1}{2}$ & $\tfrac{3}{2}^-$
&  $57.4^{+24.8}_{-21.9}$ & $4405.6$  \\
\hline
${D}^*\Sigma_c^*$ & $\tfrac{1}{2}$ & $\tfrac{1}{2}^-$
&  $19.2^{+12.7}_{-9.8}$ & $4507.8$ \\
${D}^*\Sigma_c^*$ & $\tfrac{1}{2}$ & $\tfrac{3}{2}^-$
&  $32.1^{+17.0}_{-14.2}$& $4494.9$  \\
${D}^*\Sigma_c^*$ & $\tfrac{1}{2}$ & $\tfrac{5}{2}^-$
& $61.4^{+25.9}_{-23.7}$ &$4465.6$\\
  \hline\hline
\end{tabular}
\end{table}

\section{Summary and conclusion}
\label{sec:conclusions}

The  $X_{0}(2866)$ and $X_{1}(2904)$ recently discovered by the LHCb Collaboration are the first open charm tetraquark states, which are near the $D^{\ast} \bar{K}^{\ast}$ mass threshold.  Within the one-boson exchange model we systematically investigated the $D^{(\ast)}K^{(\ast)}$ system, and showed that the $X_{0}(2866)$ can be identified as a $D^{\ast}\bar{K}^{\ast}$ hadronic molecule with $I(J^{P})=0(0^{+})$. On the other hand, the molecular interpretation for the $X_{1}(2904)$ is not favored according to our study. As a byproduct, we also predicted other likely hadronic molecules with open charm number $C=2$. In the meson-meson system,  we obtained  two molecules near the mass thresholds of $DD^{\ast}$ and $D^{\ast}D^{\ast}$    as well as two molecules  near the mass thresholds of $\bar{B}\bar{B}^{\ast}$ and $\bar{B}^{\ast}\bar{B}^{\ast}$ with $I(J^{P})=0(1^{+})$.  In the meson-baryon system, 7 molecules near the mass threshold $D^{(\ast)}\Sigma_{c}^{(\ast)}$ induced by heavy quark spin symmetry were obtained.

We note that a study of the strong decays of the $X_0(2866)$ also supports the molecular interpretation~\cite{Huang:2020abcd}, to the extent that
both studies seem to point to the fact that compact tetraquark components play an non-negligible role.  In addition to more studies on the
$X_0(2866)$ and $X_1(2904)$, experimental searches for the predicted partner states, either in the hadronic molecular picture or in the
compact tetraquark picture, will be crucial to fully disclose the nature of these states.

\section{Acknowledgments}

One of us (MZL) would like to thank the Institute of Modern Physics,
Chinese Academy of Sciences for hospitality, where part of
this work was done. This work is partly supported by the National
Natural Science Foundation of China under Grants Nos. 11735003,
11975041, 11961141004, and 11961141012. It is also partly supported by the Youth
Innovation Promotion Association CAS (No. 2016367).
\bibliography{mybib}

\end{document}